\documentclass[aps,prb,10pt,twocolumn,superscriptaddress]{revtex4-1}

\usepackage{amsmath, amssymb}    					
\usepackage{graphicx}   							
\usepackage{xcolor}     							
\usepackage{hyperref}   							
\hypersetup{colorlinks=true,allcolors=blue}			
\usepackage{textcomp}								

\newcommand{\ddt}[0]{\frac{\partial}{\partial t}}
\renewcommand{\t}[1]{\textrm{#1}}
\newcommand{\nn}[0]{\nonumber\\}
\newcommand{\an}[0]{\allowdisplaybreaks\\}
\newcommand{\mbf}[1]{\mathbf{#1}}
\renewcommand{\k}[0]{\mathbf{k}}
\newcommand{\K}[0]{\mathbf{K}}
\newcommand{\0}[0]{\mathbf{0}}

\newcommand{\R}[0]{\mathbf{R}}

\newcommand{\q}[0]{\mathbf{q}}

\newcommand{\up}[0]{\uparrow}
\newcommand{\down}[0]{\downarrow}
\newcommand{\Jsd}[0]{J_{sd}}
\newcommand{\Jpd}[0]{J_{pd}}
\newcommand{\NMn}[0]{N_\t{Mn}}

\newcommand{\omMn}{\omega_{\t{Mn}}}

\newcommand{\bs}[1]{\boldsymbol{#1}}

\newcommand{\etae}{\eta_{\t{e}}}
\newcommand{\etah}{\eta_{\t{h}}}
\renewcommand{\Im}{\textrm{Im}}
\renewcommand{\Re}{\textrm{Re}}
\newcommand{\ud}{{\uparrow/\downarrow}}
\newcommand{\du}{{\downarrow/\uparrow}}

\newcommand{\ph}[1]{\phantom{#1}}

\begin{document}

\title{Phonon impact on the dynamics of resonantly excited and hot excitons in diluted magnetic semiconductors}
\author{F. Ungar}
\affiliation{Theoretische Physik III, Universit\"at Bayreuth, 95440 Bayreuth, Germany}
\author{M. Cygorek}
\affiliation{Department of Physics, University of Ottawa, Ottawa, Ontario, Canada K1N 6N5}
\author{V. M. Axt}
\affiliation{Theoretische Physik III, Universit\"at Bayreuth, 95440 Bayreuth, Germany}

\begin{abstract}

Phonons are well known to be the main mechanism for the coupling between bright and dark excitons in nonmagnetic semiconductors.
Here, we investigate diluted magnetic semiconductors where this process is in direct competition with the scattering at localized magnetic impurities.
To this end, a recently developed quantum kinetic description of the exciton spin dynamics in diluted magnetic semiconductor quantum wells is extended by the scattering 
with longitudinal acoustic phonons.
A strong phonon impact is found in the redistribution of exciton momenta on the exciton parabola that becomes especially prominent for high temperatures and exciton distributions 
further away from the exciton resonance which are optically dark.
Despite their impact on the energetic redistribution, acoustic phonons virtually do not affect the exciton spin dynamics as the exciton-impurity interaction always dominates.
Furthermore, it turns out that the exciton spin lifetime increases by roughly one order of magnitude for nonequilibrium hot exciton distributions and, in addition, 
pronounced quantum kinetic signatures in the exciton spin dynamics appearing after resonant optical excitation are drastically reduced.

\end{abstract}

\maketitle

\section{Introduction}
\label{sec:Introduction}

Optical experiments on semiconductor nanostructures often use integrated or time-resolved photoluminsecence (PL) measurements to access physical quantities of interest.
In such measurements, the exact position of excitons on the respective exciton parabola and, thus, their kinetic energies directly influences the measured signal since
only excitons with nearly vanishing center-of-mass momenta are optically active.
Redistribution mechanisms of excitons, i.e., processes that change the exciton kinetic energy, thus either indirectly affect the PL rise time or can be directly probed by 
monitoring the longitudinal-optical (LO) phonon-assisted PL \cite{Umlauff_Direct-observation_1998}.
In semiconductors, one of the most important redistribution mechanisms is the scattering with either longitudinal-acoustic (LA) or LO phonons \cite{Umlauff_Direct-observation_1998, 
Zhao_Energy-relaxation_2002, Siantidis_Dynamics-of_2001} as it is nonelastic in nature so a significant amount of kinetic energy can be exchanged between the phonon and
the carrier system.
However, it was also found that static disorder, e.g., due to impurities or surface roughness, can have an impact on the exciton distribution \cite{Zimmermann_Excitons-in_1997, 
Zhao_Coherence-Length_2002}.
Furthermore, the redistribution of excitons to higher kinetic energies can also have consequences for the spin dynamics \cite{Ungar_Quantum-kinetic_2017, Ungar_Trend-reversal_2018}.

Although one usually tries to avoid any impurities in semiconductors in order to get as pure materials as possible, purposely doping semiconductors is a versatile technique to
controllably alter intrinsic properties.
In this context, the material class of diluted magnetic semiconductors (DMSs) \cite{Dietl_Dilute-ferromagnetic_2014, Kossut_Introduction-to_2010, Furdyna_Diluted-magnetic_1988,
Furdyna_Semiconductor-and_1988}, where impurity ions with large magnetic moments such as manganese are incorporated in a standard semiconductor lattice, displays rich physics in 
many different aspects.
For example, such a material can act as a spin aligner for an electronic current in a light-emitting diode \cite{Fiederling_Injection-and_1999}, excitonic transitions in DMSs 
can be used for the purposes of spin-noise spectroscopy \cite{Cronenberger_Atomic-like_2015}, and the material class is also promising for spintronics applications 
\cite{Dietl_A-ten_2010, Ohno_A-window_2010, Zutic_Spintronics-Fundamentals_2004, Awschalom_Challenges-for_2007}.
The doping with magnetic ions also introduces a carrier-impurity exchange scattering, which often plays the dominant role in the carrier spin dynamics \cite{Kossut_Introduction-to_2010}.

In DMSs, the majority of studies regarding the ultrafast spin dynamics is conducted at very low temperatures \cite{Camilleri_Electron-and_2001, BenCheikh_Electron-spin_2013, 
Vladimirova_Dynamics-of_2008}, so phonon effects are often disregarded.
However, when dealing with optical excitations above the band gap where carriers can show effective temperatures on the order of $10^4\,$K \cite{Cywinski_Ultrafast-demagnetization_2007} 
or when considering hot excitons which are prepared by above band-gap excitation and subsequent femtosecond relaxation via the emission of LO phonons 
\cite{Chen_Exciton-spin_2003, Poweleit_Thermal-relaxation_1997, Umlauff_Direct-observation_1998}, phonon effects evidently become important.
Since LO phonon relaxation typically occurs on much faster timescales than LA phonon relaxation, many theoretical works dealing with the spin dynamics in DMSs 
focus primarily on the former mechanism \cite{Haug_Interband-Quantum_1992, Banyai_Interband-Quantum_1992, Schilp_Electron-phonon_1994, Tsitsishvili_Phonon-induced_2005}.
In contrast, in the case of resonantly excited excitons, LO phonon emission is strongly suppressed since excitons have nearly vanishing kinetic energies, thus making it impossible to emit 
an LO phonon which carries an energy of about $30\,$meV \cite{Tsitsishvili_Phonon-induced_2005} since there are no states available with lower energies.
Furthermore, at low enough temperatures below approximately $80\,$K, LO phonon absorption processes are also absent and LA phonons dominate \cite{Rudin_Temperature-dependent_1990}.

In general, DMSs are known to exhibit pronounced many-body correlation effects\cite{Ohno_Making-Nonmagnetic_1998, DiMarco_Electron-correlations_2013, Ungar_Many-body_2018} which, when 
treated beyond the usually employed Markov approximation, lead to a significant reduction of the exciton spin-transfer rate \cite{Ungar_Quantum-kinetic_2017} as well as an unexpected trend 
reversal in their dependence on an external magnetic field \cite{Ungar_Trend-reversal_2018}.
However, these findings have been obtained only at nearly vanishing temperatures, completely neglecting the phonon scattering.
Apart from a derivation of the necessary equations in the exciton representation for quantum well systems, this work provides the foundation to study the dynamics of
excitons on a quantum kinetic level at elevated temperatures.

Here, we consider the subclass of II-VI DMS quantum well nanostructures where the impurity ions are isoelectronic, i.e., they do not lead to charge doping in the system.
As a first application, we focus on DMSs without an external magnetic field and investigate the impact of finite temperatures on the exciton distribution as well as the temperature dependence 
of spin-transfer rates while varying parameters such as the impurity content.
Since phonon emission processes are highly suppressed for the narrow optically generated exciton distributions, we also consider hot excitons where phonon emission is especially expected to be much more effective.
In all cases, a particular focus is placed on the competition between the scattering of excitons at the localized impurities in DMSs compared with the phonon scattering.
To gain insights into the importance and the signatures of non-Markovian effects, results of a standard Markovian treatment of the exciton-impurity scattering coinciding with Fermi's golden 
rule are presented together with the quantum kinetic approach, where correlations between excitons and impurities are explicitly taken into account.

We find a pronounced phonon influence on the time-resolved redistribution of the exciton momenta which can already be seen for optically generated excitons and is further enhanced for 
hot exciton distributions.
Quantitatively, this influence most strikingly manifests in a pronounced increase of the kinetic energy per exciton when phonons are accounted for.
However, despite the strong impact on the exciton occupation, the exciton spin dynamics shows little to no change with temperature, indicating a clear dominance of the magnetic exchange
interaction.
Our simulations also support the previously obtained finding that quantum kinetic effects in the spin dynamics are particularly pronounced for narrow carrier distributions close to sharp 
features in the density of states \cite{Cygorek_Non-Markovian_2015}.
Here, this statement is corroborated by the observation that exciton spin-transfer rates for hot excitons are well described by a Markovian theory, which is in drastic contrast to 
resonantly excited excitons where the Markovian description strongly overestimates the decay.
Nevertheless, quantum kinetic effects prevail in the time-resolved redistribution of exciton momenta for both excitation scenarios.
It is also found that the hot excitons display significantly longer spin lifetimes than resonantly excited electron-hole pairs.

\section{Model and phonon-induced dynamics}
\label{sec:Model-and-phonon-induced-dynamics}

In this section the Hamiltonian of our model is presented and discussed.
Furthermore, we briefly discuss the derivation of quantum kinetic equations and explicitly extend them by phonon rate equations.

\subsection{Hamiltonian}
\label{subsec:Hamiltonian}

In electron-hole representation, the Hamiltonian for the description of the exciton spin dynamics in DMSs including the influence of phonons reads 
(cf. Ref.~\onlinecite{Ungar_Quantum-kinetic_2017})
\begin{subequations}
\begin{align}
\label{eq:H_0}
H_0 =&\; \sum_{l \k} E^{l}_{\k} c^\dagger_{l \k} c_{l \k} + \sum_{v \k} E^{v}_{\k} d^\dagger_{v \k} d_{v \k},
	\an
H_\t{C}^\t{eh} =&\; - \sum_{\k \k' \q} V_{\q} \sum_{l v} c^\dagger_{l \k'+\q} d^\dagger_{v \k-\q} d_{v \k} c_{l \k'},
	\an
H_\t{lm} =&\; - \sum_{l v \k} \left( \mbf E \cdot \mbf M_{l v} c^\dagger_{l \k} d^\dagger_{v -\k} + \mbf E \cdot \mbf M_{v l} d_{v -\k} c_{l \k}\right),
	\an
 \label{eq:H_m}
H_\t{m} =&\; \frac{\Jsd}{V} \sum_{\substack{I n n' \\ l l' \k \k'}} \mbf S_{n n'} \cdot \mbf s^\t{e}_{l l'} c^\dagger_{l \k} c_{l' \k'} e^{i(\k' - \k)\cdot\R_I} \! \hat{P}^{I}_{n n'}
	\nn
	&+ \frac{\Jpd}{V} \sum_{\substack{I n n' \\ v v' \k \k'}} \mbf S_{n n'} \cdot \mbf s^\t{h}_{v v'} d^\dagger_{v \k} d_{v' \k'} e^{i(\k' - \k)\cdot\R_I} \! \hat{P}^{I}_{n n'},
 	\an
 \label{eq:H_nm}
H_\t{nm} =&\; \frac{J_0^\t{e}}{V} \sum_{\substack{I l \\ \k \k'}} c^\dagger_{l \k} c_{l \k'} e^{i(\k' - \k)\cdot\R_I}
	\nn
	&+ \frac{J_0^\t{h}}{V} \sum_{\substack{I v \\ \k \k'}} d^\dagger_{v \k} d_{v \k'} e^{i(\k' - \k)\cdot\R_I},
	\an
H_\t{ph} =&\; \sum_\q \hbar\omega_\q^\t{ph} b_\q^\dagger b_\q,
	\an
H_\t{c-ph} =&\; \sum_{\q \k} \Big( \gamma_\q^\t{e} c_{\k+\q}^\dagger c_\k b_\q + {\gamma_\q^\t{e}}^* c_\k^\dagger c_{\k+\q} b_\q^\dagger
	\nn
	&+ \gamma_\q^\t{h} d_{\k+\q}^\dagger d_\k b_\q + {\gamma_\q^\t{h}}^* d_\k^\dagger d_{\k+\q} b_\q^\dagger \Big).
\end{align}
\end{subequations}
The first part $H_0$ contains the carrier kinetic energies where the operator $c_{l\k}^\dagger$ ($c_{l\k}$) creates (annihilates) an electron in the $l$th conduction band
with wave vector $\k$ and $d_{v\k}^\dagger$ ($d_{v\k}$) is the analogous operator for holes in the $v$th valence band.
The direct Coulomb interaction between electrons and holes is given by $H_\t{C}$ with $V_q = \frac{e^2}{\epsilon_0 \epsilon q}$, where $e$ is the elementary charge,
$\epsilon_0$ denotes the vacuum permittivity, and $\epsilon$ is the static dielectric constant.
Note that the direct Coulomb interaction is actually comprised of three terms, namely the electron-electron, hole-hole, and electron-hole interaction.
However, since it turns out that only the latter yields a finite contribution in the equations of motion up to third order in the driving field \cite{Axt_A-dynamics_1994}, 
we only write down the relevant electron-hole scattering here.
The light-matter coupling is given by $H_\t{lm}$, where $\mbf E$ denotes the electric field and $\mbf M_{l v}$ is the transition dipole.

In DMSs, the dominant contribution to the spin dynamics is typically given by the magnetic $s$-$d$ and $p$-$d$ exchange interactions \cite{Kossut_Introduction-to_2010} for 
conduction band electrons and valence band holes, respectively, which are subsumed in $H_\t{m}$.
There, $\Jsd$ and $\Jpd$ are the respective coupling constants and $V$ is the volume of the semiconductor.
The indices of the coupling constants refer to the interaction of $s$-like conduction-band electrons or $p$-like holes in the valence band with the bound electrons in the $d$ shell
of the magnetic impurities, respectively.
The vector of electron (hole) spin matrices is given by $\mbf s^\t{e}_{l l'} = \bs\sigma_{l l'}$ ($s^\t{h}_{v v'} = \mbf J_{v v'}$) with the vector of Pauli matrices
$\bs\sigma_{l l'}$ and the vector of angular momentum matrices $\mbf J_{v v'}$, where $v,v' \in \{-3/2, -1/2, 1/2, 3/2\}$.
Pauli matrices are used since they provide a convenient basis for the space spanned by the spin-up and spin-down states for the conduction band.
Concerning the valence band, the angular momentum matrices are such that the quantum numbers $-3/2$ and $3/2$ correspond to the heavy-hole (hh) states, whereas the quantum numbers
$-1/2$ and $1/2$ refer to the light-hole (lh) states.
In typical semiconductor quantum wells, confinement and strain causes the hole states to split such that the lower-energy states are comprised of heavy holes 
\cite{Winkler_Spin-Orbit_2003}.
The impurity spin is decomposed into the vector of spin matrices $\mbf S_{n n'}$ with $n,n' \in \{-5/2, -3/2, ..., 5/2\}$ and the operator 
$\hat{P}^{I}_{n n'} = |I,n\rangle\langle I,n'|$, where $|I,n\rangle$ is the $n$th spin state of the $I$th impurity atom and $\R_I$ refers to its position in the lattice.
Using this representation allows us to straightforwardly discriminate between impurity operators evaluated at the same site and at different sites, which turns out to be
crucial to obtain the correct scaling behavior of the carrier-impurity spin exchange \cite{Thurn_Quantum-kinetic_2012}.
Due to the local band-gap mismatch that is created upon doping there also arises a nonmagnetic carrier-impurity interaction $H_\t{nm}$, which we model as a contact-like
interaction similar to $H_\t{m}$ with coupling constants $J_0^\t{e}$ and $J_0^\t{h}$ but without involving spin flips \cite{Cygorek_Influence-of_2017}.
Note that the magnetic and nonmagnetic scattering contributions to the Hamiltonian given by Eqs.~\eqref{eq:H_m} and \eqref{eq:H_nm} do not conserve the carrier
momentum, as should be the case for DMSs with few randomly oriented impurities.

In order to investigate the temperature dependence of the exciton spin relaxation, the model developed in Ref.~\onlinecite{Ungar_Quantum-kinetic_2017} is extended to account for
carrier-phonon scattering.
The phonons are described by $H_\t{ph}$ with creation (annihilation) operators $b_q^\dagger$ ($b_q$) for phonons with energy $\hbar\omega_\q^\t{ph}$, where $\q$ contains
the phonon momentum as well as the branch number.
The interaction with electrons (holes) is modeled by $H_\t{c-ph}$ with the coupling constant $\gamma_\q^\t{e}$ ($\gamma_\q^\t{h}$).
We use bulk phonon modes due to the relatively small change of the lattice constant with impurity content for the small doping fractions typically found in DMSs
\cite{Furdyna_Diluted-magnetic_1988} and limit the description to LA phonons, which have been found to dominate exciton line widths in semiconductors 
below $80\,$K \cite{Rudin_Temperature-dependent_1990}.
Furthermore, when focusing on excitations below the band gap near the exciton ground state, the polar piezoelectric scattering is reduced since excitons are neutral
quasiparticles and deformation potential coupling dominates \cite{Siantidis_Dynamics-of_2001}.
The corresponding coupling constants thus read
\begin{align}
\gamma_{\q,q_z}^\t{e,h} &= \sqrt{\frac{q \hbar}{2\rho V v}} D_\t{e,h}
\end{align}
for a semiconductor with density $\rho$, longitudinal sound velocity $v$, and deformation potential constants $D_\t{e,h}$ for the conduction and the valence band, respectively.
Since we only consider excitons with small or even nearly vanishing center-of-mass momenta, a linear phonon dispersion $\omega_\q^\t{ph} = vq$ is assumed.

\subsection{Phonon rate equations}
\label{subsec:Phonon-rate-equations}

Quantum kinetic equations for the exciton spin dynamics based on the Hamiltonian given by Eqs.~\eqref{eq:H_0}-\eqref{eq:H_nm} have been derived in Ref.~\onlinecite{Ungar_Quantum-kinetic_2017}
in second order in the laser field within the dynamics-controlled truncation scheme \cite{Axt_A-dynamics_1994}.
Correlations between excitons and impurities are kept as explicit dynamical variables by applying a correlation expansion and the final equations are formulated in the exciton
basis, which is convenient to model excitations below the band gap.
Within this framework, the exciton variables directly related to observables such as the optical polarization or the exciton density are those not involving impurity or phonon operators.
They are given by \cite{Ungar_Quantum-kinetic_2017, Ungar_Trend-reversal_2018}
\begin{subequations}
\begin{align}
Y_{x_1 \0}^{\sigma_1} =&\; \langle {\hat{Y}}_{\sigma -\frac{3}{2} x_1 \0} \rangle,
	\\
N_{x_1 \K_1}^{\sigma_1 \sigma_2} =&\; \langle {\hat{Y}^\dagger}_{\sigma_1 -\frac{3}{2} x_1 \K_1} {\hat{Y}}_{\sigma_2 -\frac{3}{2} x_1 \K_1} \rangle,
\end{align}
\end{subequations}
where ${\hat{Y}^\dagger}_{\sigma -\frac{3}{2} x \K}$ (${\hat{Y}}_{\sigma -\frac{3}{2} x \K}$) denotes the creation (annihilation) operator for an exciton with quantum number $x$
and two-dimensional center-of-mass wave vector $\K$.
The spin state of the electron is given by $\sigma$ and the angular momentum quantum number for the holes is $-3/2$, reflecting an optical excitation with $\sigma^-$ polarization.
We focus on semiconductors with a sufficiently large hh-lh splitting such that only the electron-spin part of the exciton needs to be considered.
Neglecting the long-range electron-hole exchange because of its much smaller interaction energy compared with the magnetic scattering \cite{Ungar_Trend-reversal_2018}, one obtains
a pinned hh spin state due to the energetic penalty of an intermediate occupation of a lh state involved in a hh spin flip \cite{Uenoyama_Hole-relaxation_1990, Ferreira_Spin-flip_1991, 
Bastard_Spin-flip_1992, Crooker_Optical-spin_1997}.
All remaining impurity-assisted variables are discussed in Appendix \ref{app:A} and may be found in Ref.~\onlinecite{Ungar_Quantum-kinetic_2017}.

Note that, apart from the bright excitons with vanishing center-of-mass momenta and electrons in the spin-up state, our theory also accounts for optically dark excitons:
First, we refer to excitons with wave numbers $K > 0$ as dark excitons since they do not directly couple to light and can therefore not recombine to emit a photon.
Thus, momentum scattering from $K \approx 0$ to $K > 0$ can convert bright excitons to dark excitons.
Second, excitons that consist of an electron with spin down are optically dark even at $K = 0$ since their recombination is spin forbidden (as the hh angular momentum quantum
number is fixed to $-3/2$).

Neglecting cross terms between phonon- and impurity-assisted variables, the phonon-induced contributions in the Markov approximation to the equations of motion can be written as
\begin{widetext}
\begin{subequations}
\label{eq:phonon-contributions}
\begin{align}
\label{eq:equation-for-Y}
\ddt|_\t{ph} \; Y_{x_1 \0}^{\sigma_1} =& - \sum_{\K} \Lambda_{x_1 x_1}^{\0 \K} \Theta\big(\omega_{x_1 \K} - \omega_{\K}^\t{ph}\big) Y_{x_1 \0}^{\sigma_1} n^\t{ph}(\omega_{x_1 \K}) 
	= - \Gamma_\t{ph} Y_{x_1 \0}^{\sigma_1} \,,
\\
\label{eq:equation-for-N}
\ddt|_\t{ph} \; N_{x_1 \K_1}^{\sigma_1 \sigma_2} =&\;
	\sum_{x \K} \Lambda_{x_1 x}^{\K_1 \K} \bigg[ \Theta\big(\omega_{x \K} - \omega_{x_1 \K_1} - \omega_{\K-\K_1}^\t{ph}\big)
	\Big( N_{x \K}^{\sigma_1 \sigma_2} \big(1+n^\t{ph}(\omega_{x \K} - \omega_{x_1 \K_1})\big) - N_{x_1 \K_1}^{\sigma_1 \sigma_2} n^\t{ph}(\omega_{x \K} - \omega_{x_1 \K_1}) \Big)
	\nn
	&+ \Theta\big(\omega_{x_1 \K_1} - \omega_{x \K} - \omega_{\K_1-\K}^\t{ph}\big)
	\Big( N_{x \K}^{\sigma_1 \sigma_2} n^\t{ph}(\omega_{x_1 \K_1} - \omega_{x \K}) - N_{x_1 \K_1}^{\sigma_1 \sigma_2} \big(1+n^\t{ph}(\omega_{x_1 \K_1} - \omega_{x_1 \K})\big) \Big)\bigg].
\end{align}
\end{subequations}
\end{widetext}
Since the exciton coherences $Y_{x_1 \0}^{\sigma_1}$ are only driven at $\K = \0$, the sum in Eq.~\eqref{eq:equation-for-Y} only needs to be performed once and yields an 
effective decay rate $\Gamma_\t{ph}$ which no longer depends on $\K$.
The terms in Eq.~\eqref{eq:equation-for-N} can be classified in terms of phonon emission and absorption processes depending on whether they are proportional to 
$1+n^\t{ph}(\Delta\omega)$ or $n^\t{ph}(\Delta\omega)$, respectively, where $n^\t{ph}$ denotes the thermal phonon occupation given by $1/(\exp{(\hbar\Delta\omega/k_BT)}-1)$.
Energy conservation is ensured by the Heaviside step function $\Theta(\Delta\omega)$ and $\omega_{x \K} = E_x + \hbar^2K^2/2M$ with the exciton mass $M$ is the exciton kinetic energy 
measured with respect to the exciton ground state, i.e., where the notation is such that $E_{1s} = 0$ and $E_x > 0$ for $x \neq 1s$.
The confinement of the excitons to the quantum well plane is taken into account by projecting the coupling to the LA phonons down to the energetically lowest well states.
Finally, the phonon matrix elements can be written as
\begin{align}
\Lambda_{x_1 x_2}^{\K_1 \K_2} =  
	\frac{P_{x_1 x_2}^{\K_1 \K_2} (\omega_{x_1 \K_1}-\omega_{x_2 \K_2})^2}{\sqrt{(\omega_{x_1 \K_1}-\omega_{x_2 \K_2})^2 - v^2(\K_1-\K_2)^2}}
	\nn
	\times \left|f\Big(\frac{1}{v}\sqrt{(\omega_{x_1 \K_1}-\omega_{x_2 \K_2})^2 - v^2(\K_1-\K_2)^2}\Big)\right|^2
\end{align}
with
\begin{align}
P_{x_1 x_2}^{\K_1 \K_2} =&\; \frac{2\pi}{\hbar \rho v^3 V} \bigg( D_\t{e}^2 F_{\etah x_1 x_2}^{\etah \K_1 \K_2} + D_\t{h}^2 F_{\etae x_1 x_2}^{\etae \K_1 \K_2} 
	\nn
	&+ 2 D_\t{e} D_\t{h} F_{-\etah x_1 x_2}^{\ph{-}\etae \K_1 \K_2} \bigg).
\end{align}
Here, infinitely high quantum well barriers are assumed so the phonon form factor becomes
\begin{align}
f(q_z) = \frac{\sin\big(\frac{q d}{2}\big)}{\frac{q d}{2}}\Big[1-\Big(\frac{q d}{2\pi}\Big)^2\Big]^{-1}.
\end{align}
An explicit expression for the exciton form factors $F_{\eta_1 x_1 x_2}^{\eta_2 \K_1 \K_2}$ can be found together with the full quantum kinetic equations of motion for the 
exciton ground state in Appendix \ref{app:A}.
Regarding the scattering with the magnetic ions we focus on the low carrier-density regime compared with the density of impurities, which allows us to describe the impurity spin
density matrix by its thermal equilibrium value using the phonon temperature throughout the dynamics \cite{Cygorek_Comparison-between_2014}.

It should be noted that it is important to first transform the equations of motion due to the carrier-phonon coupling into the exciton basis before the Markov approximation is
applied.
This is because, in the Markov approximation, one actively selects the final states which are occupied in the long-time limit since the corresponding energies directly end up
in the energy-conserving delta functions (cf. Appendix in Ref.~\onlinecite{Siantidis_Dynamics-of_2001}).
Thus, when below band-gap excitations are considered so excitons rather than quasi-free carriers are excited, a transformation to the exciton basis before introducing the
Markov approximation ensures that the correct correlated pair energies appear in Eqs.~\eqref{eq:phonon-contributions}.

Finally, in order to identify quantum kinetic signatures in the exciton dynamics, we also discuss the Markov limit of the exciton-impurity scattering \cite{Ungar_Quantum-kinetic_2017}.
In this approximation, all quantum kinetic effects are removed and only Markovian scattering processes without any memory remain.
The corresponding equations can be found in Appendix \ref{app:B}.

\section{Resonant excitation}
\label{sec:Resonant-excitation}

In the following, the phonon impact on the exciton dynamics is investigated numerically for Zn$_{1-x}$Mn$_x$Se quantum wells.
Special emphasis is put on a comparison of results where all scattering mechanisms are treated as Markovian processes without memory and calculations
where the Markov approximation is only employed for the phonon scattering, but the exciton-impurity exchange interaction is accounted for on a quantum kinetic level
so many-body effects beyond a Markovian theory are captured.
Although quantum kinetic effects have also been studied for the carrier-phonon interaction in semiconductor nanostructures, most studies have been performed for LO phonons and
the resulting carrier dynamics is typically found to be close to the results of a rate-equation approach \cite{Schilp_Electron-phonon_1994, Haug_Interband-Quantum_1992, 
Banyai_Interband-Quantum_1992}.
An indicator for the importance of quantum kinetic effects is the amount of correlation energy in the system, which is negligible for LA phonons compared with the exciton-impurity 
correlation energy \cite{Ungar_Many-body_2018}.
Furthermore, while a treatment beyond the Markov limit is required to capture an energetic redistribution of excitons due to the exchange interaction, the scattering with phonons 
involves such a redistribution already on the Markov level.
All in all, it follows that a quantum kinetic treatment is more important for the exciton-impurity interaction than for the exciton-phonon scattering, so a purely Markovian 
description of the latter can be expected to suffice.

\subsection{Occupation of the exciton parabola}
\label{subsec:Occupation-of-the-exciton-parabola}

To find out to what extent phonons can be expected to impact the dynamics of excitons, we perform simulations for a $15$-nm-wide Zn$_{1-x}$Mn$_x$Se quantum well under resonant 
optical excitation of the $1s$-hh exciton.
Except for studies where the doping concentration is varied explicitly, we focus on samples with $x = 2.5\%$.
To minimize the impact of the laser, the pulse length is chosen to be $100\,$fs and is thus short compared with typical timescales of the dynamics \cite{Thurn_Quantum-kinetic_2012, 
Ungar_Ultrafast-spin_2015, Ungar_Quantum-kinetic_2017, Ungar_Trend-reversal_2018, BenCheikh_Electron-spin_2013, Crooker_Terahertz-Spin_1996, Awschalom_Spin-dynamics_1999}.
The remaining relevant parameters used for the numerical simulations are collected in Table~\ref{tab:parameters}.

\begin{table}[h!]
	\begin{tabular}{cc}
	\hline\hline
	parameter & value for Zn$_{1-x}$Mn$_x$Se\\
	\hline
	$a\;(\t{nm})$\cite{Furdyna_Diluted-magnetic_1988} & $0.567$\\
	$m_\t{e}/m_0$\cite{Astakhov_Binding-energy_2002, Triboulet_CdTe-and_2009} & $0.15$ \\
	$m_\t{hh}/m_0$\cite{Astakhov_Binding-energy_2002, Triboulet_CdTe-and_2009} & $0.8$ \\
	$\Jsd\;(\t{meV}\,\t{nm}^3)$\cite{Furdyna_Diluted-magnetic_1988} & $-12$ \\
	$\Jpd\;(\t{meV}\,\t{nm}^3)$\cite{Furdyna_Diluted-magnetic_1988} & $50$ \\
	$J_0^\t{e}\;(\t{meV}\,\t{nm}^3)$\cite{Ungar_Quantum-kinetic_2017} & $22$ \\
	$J_0^\t{h}\;(\t{meV}\,\t{nm}^3)$\cite{Ungar_Quantum-kinetic_2017} & $0$ \\
	$\epsilon$\cite{Strzalkowski_Dielectric-constant_1976} & $9$ \\
	$D_\t{e}\;(\t{eV})$\cite{Cardona_Acoustic-deformation_1987} & $-7.4$\\
	$D_\t{h}\;(\t{eV})$\cite{Cardona_Acoustic-deformation_1987} & $-0.7$\\
	$\rho\;(\t{g cm}^{-3})$\cite{Landolt_ZnSe_1999} & $5.28$\\
	$v\;(\t{km s}^{-1})$\cite{Landolt_ZnSe_1999} & $4.21$\\
	\hline\hline
	\end{tabular}
	\caption{Selected material parameters of Zn$_{1-x}$Mn$_x$Se.
	The cubic lattice constant is denoted by $a$ and $m_0$ is the free electron mass.
	\label{tab:parameters}}
\end{table}

\begin{figure}
\centering
\includegraphics{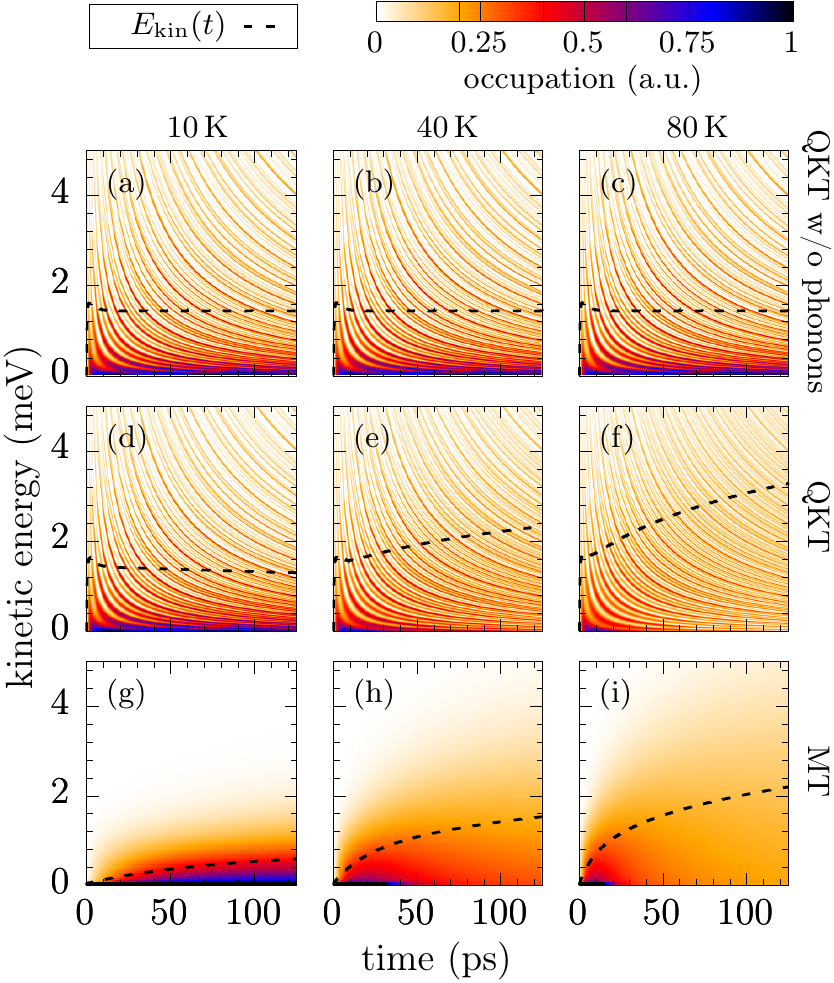}
\caption{Time- and energy-resolved occupation of the $1s$ exciton parabola obtained by the quantum kinetic theory (QKT) without accounting for phonons [(a)-(c)], the quantum kinetic
		 theory including LA phonon scattering [(d)-(f)], and the Markovian theory applied to all scattering processes [MT, (g)-(i)].
		 Apart from the occupation, we also show the time evolution of the kinetic energy per exciton $E_\t{kin}$ for each case (dashed line).
		 Results are shown for three different temperatures $T$ as indicated in the figure.}
\label{fig:occup}
\end{figure}

It is well-known that the redistribution of exciton kinetic energies due to phonon scattering leaves a fingerprint in time-resolved studies of the occupation on the $1s$ exciton parabola, 
which is also accessible in experiments \cite{Umlauff_Direct-observation_1998, Zhao_Coherence-Length_2002, Zhao_Energy-relaxation_2002}.
To investigate this impact, Fig.~\ref{fig:occup} shows the time- and energy-resolved occupation of the exciton ground state obtained by the quantum kinetic calculation with and without
phonons as well as its Markov limit.

Using the Markov approximation strictly enforces energy conservation where, however, interaction energies are completely disregarded.
This implies that the exciton-impurity scattering in the Markov approximation is unable to change the kinetic energy of excitons since it is a completely elastic process.
This can be directly seen in Eq.~\eqref{eq:Markov-limit}, where the frequency of an exciton remains unchanged during the spin-flip process.
Thus, any finite occupation of exciton kinetic energies in the Markovian theory seen in the bottom row of Fig.~\ref{fig:occup} is purely due to phonons.
The phonon influence increases with increasing temperature because phonon emission is prohibited for optically generated excitons so phonon absorption processes are
required to affect the exciton occupation.
Figure~\ref{fig:occup} also reveals that not only are higher kinetic energies reached at elevated temperatures, the scattering also becomes noticeably faster.
While for $10\,$K only states below $2\,$meV are occupied after more than $100\,$ps, at $80\,$K exciton kinetic energies exceed $4\,$meV after approximately $50\,$ps.

Previous works have shown that, when treating the exciton-impurity scattering on a quantum kinetic level, one obtains a significant redistribution of exciton kinetic 
energies which is accompanied by a build-up of a many-body correlation energy \cite{Thurn_Non-Markovian_2013, Ungar_Quantum-kinetic_2017, Ungar_Trend-reversal_2018, Ungar_Many-body_2018}.
In the top row of Fig.~\ref{fig:occup}, where the interaction with phonons is completely switched off so only the exchange interaction remains, this manifests in an occupation of 
exciton states with high kinetic energies up to $4\,$meV and above.
Note that the higher-energy states are not uniformly occupied; 
rather, a typical pattern appears which is caused by the memory kernel due to the exciton-impurity interaction.
Since this memory kernel is proportional to $\frac{\sin(\omega t)}{\omega}$, it is large for short times and small energies and shows a damped oscillation for larger 
times \cite{Ungar_Quantum-kinetic_2017}.
This is a manifestation of the energy-time uncertainty which allows the violation of strict single-particle energy conservation rules on short timescales.
Here, this effect is combined with a relaxation of the system to a new energy eigenstate that forms as a result of the interaction between excitons and impurities when 
correlations between them are accounted for.
This is why elevated exciton kinetic energies remain occupied even in the long-time limit.

Comparing the top row in Fig.~\ref{fig:occup} with the bottom row, it becomes clear that the quantum kinetic redistribution is much stronger as well as faster than the phonon-induced 
scattering, which can be seen from the fact that after only a few picoseconds energies up to $4\,$meV and above are occupied even at a low temperature of $10\,$K.
However, when comparing the quantum kinetic calculations for different temperatures, one can still clearly see the phonon influence by looking at excitons at very low energies.
There, the occupation visibly decreases with time just like in the Markovian case, suggesting that phonons cause a smoothing out of the overall exciton occupation, which is still 
peaked near $E = 0$ in the quantum kinetic case at $10\,$K.
All in all, phonons thus cause a more efficient coupling of excitons with vanishing center-of-mass motion towards the optically dark states away from $K = 0$, especially at
elevated temperatures.

The phonon influence on the exciton occupation can also be studied on a more quantitative level by looking at the kinetic energy per exciton, which is indicated in Fig.~\ref{fig:occup} 
by a dashed line.
As mentioned before, when the scattering of excitons with impurities is described on the Markovian level, there is no way for excitons to change their kinetic energy
after the optical pulse except via the emission or absorption of phonons.
Thus, the increase of the kinetic energy per exciton observed in Figs.~\ref{fig:occup}(g)-(i) is exclusively due to phonon scattering and consistently increases with rising
temperature.
To be specific, phonon absorption processes cause the excitons to reach energies in excess of $2\,$meV on the order of $100\,$ps at $80\,$K.
The kinetic energy per exciton also directly reflects the many-body correlation energy per exciton that is built up after the optical excitation due to the non-Markovian nature
of the exciton-impurity exchange interaction \cite{Ungar_Quantum-kinetic_2017}.
From Figs.~\ref{fig:occup}(a)-(c), a kinetic energy per exciton of about $1.5\,$meV can be determined for the parameters considered here, which is significantly larger than the
phonon contribution for low temperatures and only becomes smaller than the phonon-induced kinetic energy at temperatures exceeding a few $10\,$K.
As discussed previously in terms of the redistribution of excitons towards the optically dark states, the phonon impact is also clearly visible in the increase of the kinetic energy 
per exciton in Figs.~\ref{fig:occup}(d)-(f).
At a temperature of $80\,$K, the combined influence of the many-body correlations and the phonon scattering even cause energies in excess of $3\,$meV.

\subsection{Exciton spin-transfer rates}
\label{subsec:Exciton-spin-transfer-rates}

\begin{figure}
\centering
\includegraphics{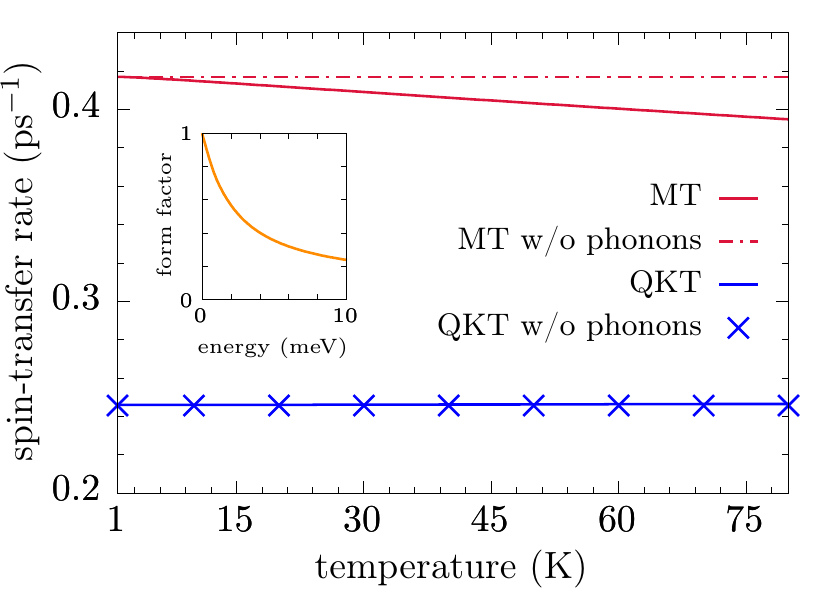}
\caption{Exciton spin-transfer rate as a function of temperature obtained by the Markovian theory (MT) and the quantum kinetic theory (QKT) after resonant excitation of 
		 the $1s$ exciton.
		 We compare simulations with and without (w/o) phonons.
		 The inset depicts the diagonal of the exciton form factor $F_{\etah 1s 1s}^{\etah \omega \omega}$ appearing in Fermi's golden rule as a function of energy $E = \hbar\omega$.}
\label{fig:rate_vs_T}
\end{figure}

Although phonons do not introduce spin-flip transitions, at least not without some type of spin-orbit coupling, they do cause a significant change in the kinetic energy of excitons 
by either phonon absorption or emission processes.
Thus, it is an interesting question to ask whether these processes also affect the spin dynamics of excitons in DMSs, for which the magnetic exchange interaction is commonly
regarded as the most important process \cite{Kossut_Introduction-to_2010}.
As a measure for the spin dynamics, Fig.~\ref{fig:rate_vs_T} shows the spin-transfer rate for the exciton-bound electron as a function of temperature.
Since the spin dynamics is nonexponential in general \cite{Ungar_Quantum-kinetic_2017, Ungar_Trend-reversal_2018}, the rate is extracted as the inverse time that it takes the spin to 
decay to $\frac{1}{e}$ of its maximum value after the optical pulse.

Focusing first on the results when the exciton-impurity scattering is treated as a Markovian process, we find that phonons cause a slowdown of the spin-transfer rate, which becomes 
more significant with rising temperature.
Whereas the phonon influence is completely negligible for temperatures below $4\,$K, at a temperature of $80\,$K phonons cause a change in the rate of approximately $5\,$\%.
This behavior is directly connected to the redistribution of exciton momenta caused by the interaction with LA phonons. 
Remembering that we are considering optically created excitons with quasivanishing wave vector, it is clear that phonons have almost no impact on the spin dynamics for very low 
temperatures since, in this case, phonon absorption processes are highly suppressed.
Such processes are, however, required to change the exciton kinetic energy, as phonon emission is prohibited for optically created excitons because there are no states available with 
lower energies to scatter to.
As the temperature increases, phonon absorption processes become increasingly probable.
Since the exciton-phonon interaction is spin conserving and phonon emission is suppressed, the scattering of an exciton by a phonon thus increases its kinetic energy on average.

The observed slowdown of the spin-transfer rate with increasing temperature in Fig.~\ref{fig:rate_vs_T} for the Markov approximation is directly related to the previously introduced 
exciton form factor $F_{\etah 1s 1s}^{\etah K_1 K_2}$, whose diagonal is one for $K = 0$ and quickly decreases for larger center-of-mass momenta, i.e., larger kinetic energies 
\cite{Ungar_Quantum-kinetic_2017, Bastard_Spin-flip_1992}.
Since the exciton form factor enters the rate obtained in Fermi's golden rule [cf. also Eq.~\eqref{eq:Markov-limit}] which, for an exciton with kinetic energy $\hbar\omega$, is given 
by \cite{Ungar_Quantum-kinetic_2017}
\begin{align}
\label{eq Markov spin-transfer rate}
\tau^{-1}_{\omega} &= \frac{35}{12} \frac{\NMn I \Jsd^2 M}{\hbar^3 d V} F_{\etah 1s 1s}^{\etah \omega \omega},
\end{align}
the rate becomes smaller when the form factor is evaluated at larger energies.
In the above notation, the center-of-mass momentum of an exciton is connected with its frequency via the usual relation $\omega = \frac{\hbar K^2}{2M}$.
The decrease of the diagonal elements of the exciton form factor are shown in the inset in Fig.~\ref{fig:rate_vs_T}.
Thus, the increasing influence of phonons on the Markovian spin-transfer rate with rising temperature follows from the increase of the kinetic energy per exciton observed in
Figs.~\ref{fig:occup}(g)-(i).
On the other hand, when the quantum kinetic theory is used to calculate the rates, the phonon influence is more or less absent since there is already a significant kinetic energy
per exciton even without phonons [cf. Figs.~\ref{fig:occup}(a)-(c)].

\begin{figure}
\centering
\includegraphics{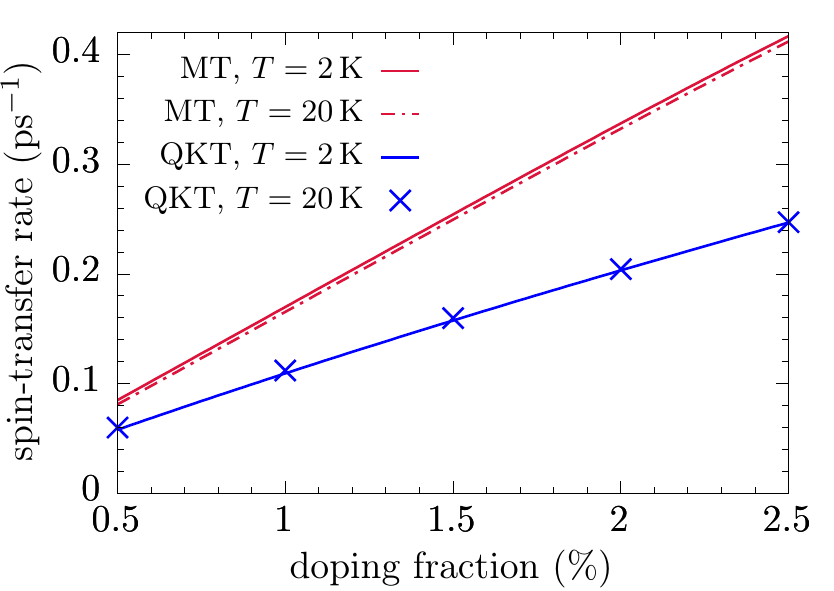}
\caption{Exciton spin-transfer rate as a function of Mn doping fraction obtained by the Markovian theory (MT) and the quantum kinetic theory (QKT) after resonant excitation of 
		 the $1s$ exciton.
		 The simulations are performed for two different temperatures $T$.}
\label{fig:rate_vs_xMn}
\end{figure}

Turning to the quantum kinetic results in Fig.~\ref{fig:rate_vs_T}, one first of all notices a generally lower spin-transfer rate compared with the Markovian results, an effect which 
is due to the abrupt disappearance of the density of states at the onset of the exciton parabola, which effectively cuts off the memory kernel appearing in the quantum kinetic treatment 
of the exciton-impurity interaction \cite{Ungar_Quantum-kinetic_2017, Ungar_Trend-reversal_2018}.
Concerning the phonon influence, however, the spin-transfer rate is entirely dominated by the exchange interaction and remains constant over the whole temperature range considered here.
Surprisingly, even though phonons increase the redistribution of excitons towards optically dark states, their effect on the spin dynamics is marginal.
It should be noted here that this behavior is expected to change drastically when optical phonon scattering becomes important, which is either at higher temperatures or when looking 
at hot excitons with kinetic energies above the LO phonon threshold \cite{Siantidis_Dynamics-of_2001, Tsitsishvili_Magnetic-field_2006, Chen_Exciton-spin_2003, 
Poweleit_Thermal-relaxation_1997, Umlauff_Direct-observation_1998}.

The observation that the exchange scattering dominates over the phonon scattering remains valid also for lower doping fractions, as can be seen in Fig.~\ref{fig:rate_vs_xMn} where 
the spin-transfer rate in the Markov approximation and using the quantum kinetic theory is depicted as a function of the impurity concentration for two different temperatures.
Again, when the exciton-impurity scattering is treated as a Markovian process, a slight phonon influence is visible when going from $2\,$K to $20\,$K.
In contrast, the quantum kinetic result remains virtually unaffected by the increase in temperature.

\section{Hot excitons}
\label{sec:Hot-excitons}

Having discussed the phonon impact on resonantly excited excitons, we now turn to an initially nonequilibrium exciton distribution, also referred to as \emph{hot excitons}, 
that can be generated, e.g., by optical excitation above the band gap and subsequent formation of excitons on the $1s$ parabola via fast LO phonon emission \cite{Umlauff_Direct-observation_1998}.
Apart from the resulting exciton spin-transfer rates, the dynamics of the exciton distribution is also investigated.

\subsection{Time evolution of the exciton distribution}
\label{subsec:Time-evolution-of-the-exciton-distribution-hot}

Since for the case of resonant optical excitation the phonon impact is most clearly visible in the time evolution of the exciton distribution, it is interesting to
ask the question whether this is also true for hot excitons.
As in the previous section, we focus on a $15$-nm-wide Zn$_{1-x}$Mn$_{x}$Se DMS quantum well at varying temperatures and consider a Mn content of $x = 2.5\%$ if the doping 
concentration is not varied explicitly.
However, instead of modeling a resonant optical excitation of the $1s$ exciton, we instead consider an initially hot exciton distribution far away from $K = 0$ as created, e.g., 
by LO phonon emission after above band-gap excitation \cite{Chen_Exciton-spin_2003, Poweleit_Thermal-relaxation_1997, Umlauff_Direct-observation_1998}.
For the numerical simulations, we assume a Gaussian distribution on the $1s$ exciton parabola centered at an energy of $10\,$meV with a standard deviation of $1\,$meV,
which translates to a distribution with a FWHM of roughly $2.5\,$meV similar to what has been observed in experiments \cite{Umlauff_Direct-observation_1998}.
The remaining parameters are the same as in Table~\ref{tab:parameters}.

\begin{figure}
\centering
\includegraphics{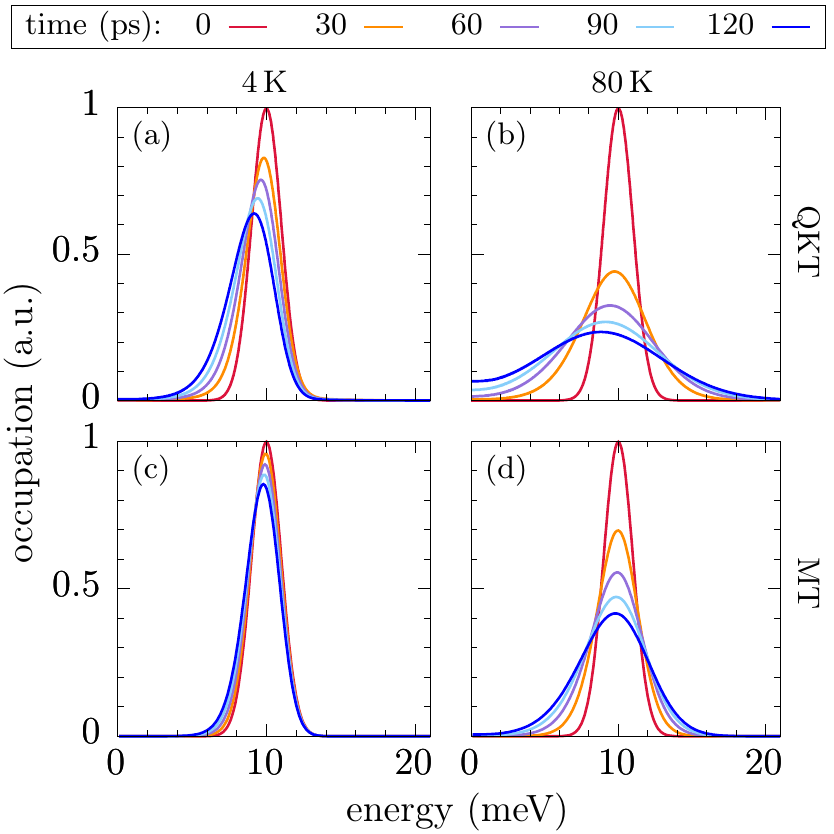}
\caption{Snapshots of the dynamics of the $1s$ exciton distribution of hot excitons.
		 The simulations show results using (a) the quantum kinetic theory (QKT) with phonons at $4\,$K and (b) at $80\,$K as well as results obtained by
		 (c) the Markovian theory (MT) with phonons at $4\,$K and (d) at $80\,$K, respectively.}
\label{fig:hot_occup}
\end{figure}

Figure~\ref{fig:hot_occup} displays several snapshots of the exciton distribution for two different temperatures as calculated by either the quantum kinetic or
the Markovian theory.
Focusing first of all on the Markovian results in Figs.~\ref{fig:hot_occup}(c) and \ref{fig:hot_occup}(d), the phonon influence clearly causes a broadening of the
exciton distribution which becomes stronger at elevated temperatures.
The asymmetry between phonon absorption and emission processes also affects the distributions since, for $T=4\,$K, only a broadening towards the low-energy side is observed
while for $80\,$K there is also a significant broadening on the high-energy side.
This is because, while phonon emission processes already occur even at very low temperatures, phonon absorption processes are proportional to the number of thermally
excited phonons and therefore only become important at higher temperatures.

Turning to the quantum kinetic results in Fig.~\ref{fig:hot_occup}(a) and Fig.~\ref{fig:hot_occup}(b), it becomes clear that the redistribution is greatly enhanced due to
the correlation energy already at very low temperatures, similar to what has been observed for resonantly excited excitons.
Furthermore, a slight shift of the maximum of the exciton distribution towards lower energies is found, which is almost completely absent in the Markovian simulations.
Figure~\ref{fig:hot_occup}(b) also shows a sizable exciton occupation of bright states near $E = 0$ already before $100\,$ps in contrast to the simulation for low temperatures 
in Fig.~\ref{fig:hot_occup}(a).
Comparing the two figures in question with the respective Markovian results, it becomes clear that a purely Markovian theory significantly underestimates the population of
bright exciton states.

\subsection{Exciton spin-transfer rates}
\label{subsec:Exciton-spin-transfer-rates-hot}

\begin{figure}
\centering
\includegraphics{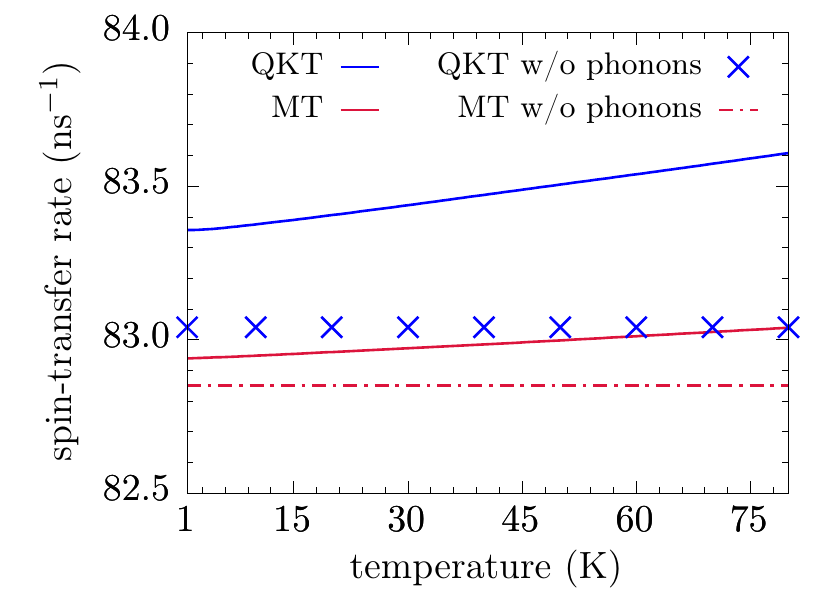}
\caption{Exciton spin-transfer rate as a function of temperature obtained by the Markovian theory (MT) and the quantum kinetic theory (QKT) assuming an initially hot 
		 exciton distribution.
		 We compare simulations with and without (w/o) phonons.}
\label{fig:rate_vs_T_hot}
\end{figure}

Looking at the spin-transfer rate of hot excitons as a function of temperature depicted in Fig.~\ref{fig:rate_vs_T_hot}, the most striking feature is that the spin relaxation is slowed down 
by an order of magnitude compared with the case of resonant optical excitation (cf. Fig.~\ref{fig:rate_vs_T}).
As discussed already in Sec.~\ref{subsec:Exciton-spin-transfer-rates-hot}, this slowdown is directly related to the decrease of the exciton form factor for larger energies.
Even though the hot exciton distribution is quite broad and thus covers a range of kinetic energies, the reduction of the form factor fits very well to the observed reduction of
the rate.
For the Markovian calculation without phonons, the rate after resonant optical excitation is approximately $\tau^{-1}_\text{res} \approx 0.417\,$ps$^{-1}$ whereas for hot excitons it is 
given by $\tau^{-1}_\text{hot} \approx 0.083\,$ps$^{-1}$, which yields a ratio of about $0.2$.
For the parameters considered here, the exciton form factor is also reduced to approximately $0.2$ when evaluated at an energy of $10\,$meV, thus quantitatively explaining the drastic reduction of the spin-transfer rate.
Since shorter decay rates translate to longer spin lifetimes, this effect can potentially be utilized to protect any information encoded in the exciton spin degree of freedom.

Apart from this striking quantitative difference with respect to the case of resonant optical excitation, the results from the quantum kinetic theory are much closer to those obtained by
the Markovian theory for hot excitons.
Note the change in units from picoseconds in Fig.~\ref{fig:rate_vs_T} to nanoseconds in Fig.~\ref{fig:rate_vs_T_hot}.
This points to a drastic reduction of quantum kinetic effects on the spin dynamics for hot excitons, a finding which is expected since previous works have shown
quantum kinetic effects to be particularly strong near sharp features in the density of states \cite{Cygorek_Non-Markovian_2015, Ungar_Quantum-kinetic_2017}.
The broad distribution of the hot excitons, which is a few meV away from the center of the exciton parabola where the constant density of states is abruptly cut off, thus largely inhibits
pronounced non-Markovian features in the spin dynamics so a Markovian approach yields very similar decay rates.

Turning finally to the phonon influence on the spin dynamics, Fig.~\ref{fig:rate_vs_T_hot} reveals a similar behavior as already found for resonant excitation in Fig.~\ref{fig:rate_vs_T},
namely that phonons only marginally influence the spin dynamics overall.
Thus, even though the impact of phonons on the dynamics of the exciton occupation is even stronger for hot excitons, the spin dynamics is once more dominated by the exchange interaction.
However, compared with Fig.~\ref{fig:rate_vs_T}, here phonons are found to slightly increase the spin-transfer rate in the quantum kinetic as well as the Markovian simulation.
In addition, instead of a slowdown of the rate, even without phonons the quantum kinetic theory predicts a slightly larger spin-transfer rate compared with the Markovian results 
in the case of hot excitons.
It can be argued that these two observations rely on a similar mechanism which, however, stems from completely different physical processes.
Including phonons in the model evidently allows for a scattering of excitons towards lower energies which becomes stronger with higher temperatures.
The resulting decrease of the kinetic energy of the scattered part of the exciton distribution then causes the spin-transfer rate to be evaluated at lower energies,
where the exciton form factor is larger, which causes an increased spin-transfer rate.
Since the quantum kinetic theory also predicts a strong redistribution of carrier momenta due to the correlation energy (cf. Fig.~\ref{fig:occup}), the quantum kinetic rate
is expected to increase for the same reason.
The strong phonon impact on the exciton distribution found for the quantum kinetic simulations in Fig.~\ref{fig:hot_occup} directly translates to the more pronounced phonon impact
in the quantum kinetic simulations in Fig.~\ref{fig:rate_vs_T_hot}.

\begin{figure}
\centering
\includegraphics{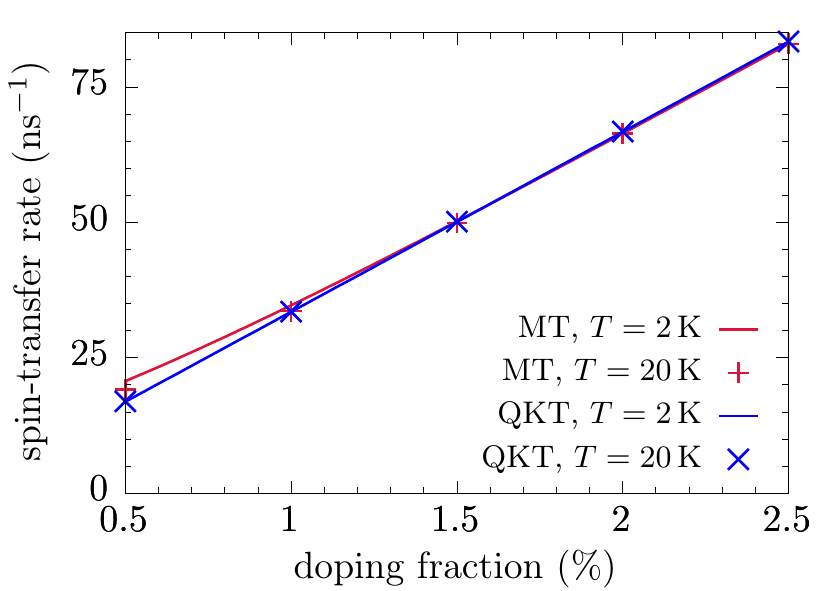}
\caption{Exciton spin-transfer rate as a function of Mn doping fraction obtained by the Markovian theory (MT) and the quantum kinetic theory (QKT) assuming an initially hot 
		 exciton distribution.
		 The simulations are performed for two different temperatures $T$.}
\label{fig:rate_vs_xMn_hot}
\end{figure}

Regarding the dependence of the spin-transfer rate on the doping fraction as plotted in Fig.~\ref{fig:rate_vs_xMn_hot}, we find only a small quantitative difference
between the quantum kinetic and the Markovian theory for small doping fractions which becomes smaller with increasing impurity content.
Similar to the case of resonant excitation, the phonon influence is again almost completely negligible for hot excitons on the scale of the figure.
However, comparing Fig.~\ref{fig:rate_vs_xMn_hot} to the the case of resonant optical excitation in Fig.~\ref{fig:rate_vs_xMn} and noting the change in units, we find much slower
rates in accordance with the exciton form factor.

\section{Conclusion}
\label{sec:Conclusion}

We have investigated the phonon impact on the exciton dynamics in DMSs.
To this end, we have extended a quantum kinetic description of the exciton spin dynamics in DMS quantum wells by accounting for the scattering with LA phonons on the Markov level.
The derived equations open up the possibility to also study dynamical processes at elevated temperatures below the LO phonon threshold.
Numerical calculations without an external magnetic field show a pronounced impact of LA phonons on the distribution of resonantly excited excitons on the $1s$ parabola, revealing that 
they introduce a significant occupation of states with higher kinetic energies in addition to the redistribution caused by quantum many-body effects due to the scattering of excitons 
with the localized impurities in DMSs.
In contrast to the inelastic phonon scattering, which always induces transitions to states with different energies, the exciton-impurity exchange interaction is an elastic process
and is therefore single-particle energy conserving when it is treated as a Markovian process.
Although a phonon impact on the time- and energy-resolved exction distribution can also be seen using the quantum kinetic theory, it is found that the quantum kinetic redistribution 
enabled by a many-body correlation energy is the dominant process.

Considering a broad initial distribution of hot excitons as, e.g., generated by above band-gap excitation and subsequent LO phonon emission, reveals a more prominent role of LA phonon 
emission since states with lower energy become available.
For these situations, the quantum kinetic redistribution is found to be greatly assisted by the phonon scattering, which leads to an increased broadening, especially on the low-energy
side.
Together, both scattering mechanisms provide an efficient pathway for the conversion of optically dark excitons to bright states.

Despite the clear influence of phonons on the exciton occupation, the phonon influence on exciton spin-transfer rates for the case of resonantly excited excitons is found to be 
completely negligible over a wide range of temperatures and doping fractions when the exchange interaction between excitons and impurities is treated on a quantum kinetic level.
If the exchange interaction is treated as a purely Markovian process without any memory, LA phonons do, in fact, increasingly inhibit the spin transfer at elevated temperatures 
since they only cause spin conserving scattering and therefore provide a competition to spin-flip interactions.

Compared to a resonant excitation scheme, hot excitons show a significantly longer spin lifetime, making them promising for the protection of information stored in the spin
degree of freedom.
This long spin lifetime is related to the decrease of the exciton form factor with increasing exciton kinetic energy, thus causing significantly smaller spin-transfer rates.
For these situations, quantum kinetic effects are strongly reduced so a Markovian treatment of the spin transfer can be justified.
Finally, phonons cause an increase of the spin-transfer rate for hot excitons in both considered theoretical models, albeit on a rather small scale.

\section*{Acknowledgment}
\label{sec:Acknowledgment}

We gratefully acknowledge the financial support of the Deutsche Forschungsgemeinschaft (DFG) through Grant No. AX17/10-1.

\appendix
\section{Quantum kinetic equations of motion for the exciton ground state}
\label{app:A}

In Ref.~\onlinecite{Ungar_Trend-reversal_2018}, a quantum kinetic description of the exciton spin dynamics without phonon scattering has been derived in the form of coupled equations
for excitonic variables.
Focusing on the exciton ground state and performing an average over angles in reciprocal space, the necessary variables are the exciton density $n_{K_1}$, the $l$th exciton spin
component $s_{K_1}^l$, and the coherences $y^\ud$ for the different spin states.
The correlations of excitons with magnetic impurities are explicitly taken into account via the variables $Q_{\eta l K_1}^{\ph{\eta} \alpha K_2}$, nonmagnetic correlations are
given by $Z_{\eta \ph{\alpha} K_1}^{\ph{\eta} \alpha K_2}$.
Similarly, correlations between the coherences and impurities are denoted by $q_{\eta l K_1}^{\ph{\eta} \ud}$ and $z_{\eta K_1}^{\ph{\eta} \ud}$ for the magnetic and nonmagnetic
interactions, respectively.
The notation is chosen such that $l \in \{1,2,3\}$, $\alpha \in \{0,1,2,3\}$, and $\eta \in \{\etae,\etah\}$.
Explicit expressions regarding the definition of the all the above variables can be found in Ref.~\onlinecite{Ungar_Quantum-kinetic_2017}.

On a more technical level, we note that obtaining the equations of motion includes the assumption of a system which is homogeneous on average, so the positions of impurities 
no longer appear explicitly in the variables.
Nevertheless, as discussed in detail in Ref.~\onlinecite{Thurn_Quantum-kinetic_2012}, our model still accounts for the transfer of occupations between states with different
center-of-mass momenta.
The reason for this is that, instead of performing the averaging procedure on the level of the Hamiltonian and thereby artificially enforcing momentum conservation, we only
carry out the averaging over impurity positions on the level of the equations of motion.
The fact that our theory does indeed allow a change of exciton momenta can be directly seen from Fig.~\ref{fig:occup}, where the kinetic energy (and, thus, the momenta) of excitons
remains finite even at long times.
Regarding the variables of our model, momentum non-conservation during scattering processes with impurities is accounted for by the magnetic and nonmagnetic correlations.

If the coupling to LA phonons is treated on the Markovian level and cross terms between different interactions are disregarded, the quantum kinetic equations of motion become modified
but no new dynamical variables have to be introduced.
The modified equations including the phonon scattering read
\begin{widetext}
\begin{subequations}
\label{eqs:EOM-for-summed-variables}
\begin{align}
\label{eq:EOM-for-summed-variables-n}
\ddt n_{K_1} =&\;
	\frac{2}{\hbar}\mbf E \cdot \mbf M \Im\big[y^\uparrow \phi_{1s} \big] \delta_{K_1,0}
	- \frac{\Jsd \NMn}{\hbar V^2} \sum_{l K} 2\Im\big[ Q_{-\etah l K}^{\ph{-\etah} l K_1} \big]
	+ \frac{\Jpd \NMn}{\hbar V^2} \sum_{K} \Im\big[ Q_{\etae z K}^{\ph{\etae} 0 K_1} \big]
	\nn
	&- \frac{J_0^\t{e} \NMn}{\hbar V^2} \sum_{K} 2\Im\big[ Z_{-\etah \ph{0} K}^{\ph{-\etah} 0 K_1} \big]
	- \frac{J_0^\t{h} \NMn}{\hbar V^2} \sum_{K} 2\Im\big[ Z_{\etae \ph{0} K}^{\ph{\etae} 0 K_1} \big]
	\nn
	&+ \sum_{K} \Lambda_{1s 1s}^{K_1 K} \bigg[ \Theta\big(\omega_{K} - \omega_{K_1} - \omega_{K-K_1}^\t{ph}\big)
	\Big( n_{K} \big(1+n^\t{ph}(\omega_{K} - \omega_{K_1})\big) - n_{K_1} n^\t{ph}(\omega_{K} - \omega_{K_1}) \Big)
	\nn
	&+ \Theta\big(\omega_{K_1} - \omega_{K} - \omega_{K_1-K}^\t{ph}\big)
	\Big( n_{K} n^\t{ph}(\omega_{K_1} - \omega_{K}) - n_{K_1} \big(1+n^\t{ph}(\omega_{K_1} - \omega_{K})\big) \Big)\bigg],
	\an
\label{eq:EOM-for-summed-variables-s}
\ddt s_{K_1}^l =&\;
	\frac{1}{\hbar}\mbf E \cdot \mbf M \Big( \Im\big[ y^\uparrow \phi_{1s} \big] \delta_{K_1,0}\delta_{l,z}
	+ \Im\big[ y^\downarrow \phi_{1s} \big] \delta_{K_1,0}\delta_{l,x}
	- \Re\big[ y^\downarrow \phi_{1s} \big] \delta_{K_1,0}\delta_{l,y} \Big)
	\nn
	&+ \frac{\Jsd \NMn}{\hbar V^2} \sum_{K} \Big( \sum_{j k} \epsilon_{jkl} \Re\big[Q_{-\etah j K}^{\ph{-\etah} k K_1}\big] 
	- \frac{1}{2} \Im\big[Q_{-\etah l K}^{\ph{-\etah} 0 K_1}\big] \Big)
	+ \frac{\Jpd \NMn}{\hbar V^2} \sum_{K} \Im\big[ Q_{\etae z K}^{\ph{\etae} l K_1} \big] 
	\nn
	&- \frac{J_0^\t{e} \NMn}{\hbar V^2} \sum_{K} 2\Im\big[ Z_{-\etah \ph{l} K}^{\ph{-\etah} l K_1} \big]
	- \frac{J_0^\t{h} \NMn}{\hbar V^2} \sum_{K} 2\Im\big[ Z_{\etae \ph{l} K}^{\ph{\etae} l K_1} \big]
	\nn
	&+ \sum_{K} \Lambda_{1s 1s}^{K_1 K} \bigg[ \Theta\big(\omega_{K} - \omega_{K_1} - \omega_{K-K_1}^\t{ph}\big)
	\Big( s_{K}^l \big(1+n^\t{ph}(\omega_{K} - \omega_{K_1})\big) - s_{K_1}^l n^\t{ph}(\omega_{K} - \omega_{K_1}) \Big)
	\nn
	&+ \Theta\big(\omega_{K_1} - \omega_{K} - \omega_{K_1-K}^\t{ph}\big)
	\Big( s_{K}^l n^\t{ph}(\omega_{K_1} - \omega_{K}) - s_{K_1}^l \big(1+n^\t{ph}(\omega_{K_1} - \omega_{K})\big) \Big)\bigg],
	\an
\label{eq:EOM-for-summed-variables-y}
\ddt y^\ud =&\;
	\frac{i}{\hbar}\mbf E \cdot \mbf M \phi_{1s} \delta_{\ud, \uparrow}
	-i \Big( \omega_{0} + \frac{(J_0^\t{e}+J_0^\t{h}) \NMn}{\hbar V} \Big) y^\ud
	- i \frac{\Jsd \NMn}{2\hbar V^2} \sum_{K} \Big(\pm q_{-\etah z K}^{\ph{-\etah} \ud} + q_{-\etah \mp K}^{\ph{-\etah} \du}\Big)
	\nn
	&+ i \frac{\Jpd \NMn}{2\hbar V^2} \sum_{K} q_{\etae z K}^{\ph{\etae} \ud}
	- i \frac{J_0^\t{e} \NMn}{\hbar V^2} \sum_{K} z_{-\etah K}^{\ph{-\etah} \ud}
	- i \frac{J_0^\t{h} \NMn}{\hbar V^2} \sum_{K} z_{\etae K}^{\ph{\etae} \ud}
	\nn
	&- \sum_{K} \Lambda_{1s 1s}^{0 K} \Theta\big(\omega_{K} - \omega_{K}^\t{ph}\big) n^\t{ph}(\omega_{K}) y^\ud,
	\an
\ddt q_{\eta l K_1}^{\ph{\eta} \ud} =&
	- i \Big( \omega_{K_1} + \frac{I (J_0^\t{e}+J_0^\t{h}) \NMn}{\hbar V} \Big) q_{\eta l K_1}^{\ph{\eta} \ud}
	- i \frac{I \Jsd}{2\hbar} F_{\ph{-}\eta\ph{_h} 1s 1s}^{-\etah 0 K_1} \Big(\pm\langle S^lS^z \rangle y^\ud + \langle S^lS^\mp \rangle y^\du\Big)
	\nn
	&+ i \frac{I \Jpd}{2\hbar} \langle S^lS^z \rangle F_{\eta\ph{_e} 1s 1s}^{\etae 0 K_1} y^\ud,
	\an
\ddt z_{\eta K_1}^{\ph{\eta} \ud} =&
	- i \Big( \omega_{K_1} + \frac{I (J_0^\t{e}+J_0^\t{h}) \NMn}{\hbar V} \Big) z_{\eta K_1}^{\ph{\eta} \ud}
	- i \frac{I}{\hbar} \Big( J_0^\t{e} F_{\ph{-}\eta\ph{_h} 1s 1s}^{-\etah 0 K_1} + J_0^\t{h} F_{\eta\ph{_e} 1s 1s}^{\etae 0 K_1} \Big) y^\ud,
	\an
\label{eq:EOM-for-summed-variables-Q1}
\ddt Q_{\eta l K_1}^{\ph{\eta} 0 K_2} =&
	- i \big(\omega_{K_2} - \omega_{K_1}\big) Q_{\eta l K_1}^{\ph{\eta} 0 K_2}
	+ \sum_{j k} \epsilon_{jkl} \omMn^j Q_{\eta k K_1}^{\ph{\eta} 0 K_2}
	+ \frac{i}{2\hbar}\mbf E \cdot \mbf M \Big(\big(q_{\eta l K_1}^{\ph{\eta} \uparrow} \phi_{1s}\big)^* \delta_{K_2,0} 
	- q_{\eta l K_2}^{\ph{\eta} \uparrow} \phi_{1s}\delta_{K_1,0}\Big)
	\nn
	&+ i \frac{I \Jsd}{\hbar} F_{\ph{-}\eta\ph{_h} 1s 1s}^{-\etah K_1 K_2} \sum_j \Big( \langle S^jS^l \rangle s_{K_2}^j - \langle S^lS^j \rangle s_{K_1}^j \Big)
	- i \frac{I \Jpd}{\hbar} F_{\eta\ph{_e} 1s 1s}^{\etae K_1 K_2} \frac{1}{2} \Big( \langle S^zS^l \rangle n_{K_2} - \langle S^lS^z \rangle n_{K_1} \Big),
	\an
\label{eq:EOM-for-summed-variables-Q2}
\ddt Q_{\eta l K_1}^{\ph{\eta} m K_2} =&
	- i \big(\omega_{K_2} - \omega_{K_1}\big) Q_{\eta l K_1}^{\ph{\eta} m K_2}
	\nn
	&+ \frac{i}{2\hbar}\mbf E \cdot \mbf M \bigg[ \Big(\big(q_{\eta l K_1}^{\ph{\eta} \uparrow} \phi_{1s}\big)^* \delta_{K_2,0} 
	\! - \! q_{\eta l K_2}^{\ph{\eta} \uparrow} \phi_{1s}\delta_{K_1,0}\Big) \delta_{m,z}
	+ \Big(\big(q_{\eta l K_1}^{\ph{\eta} \downarrow} \phi_{1s}\big)^* \delta_{K_2,0} 
	\! - \! q_{\eta l K_2}^{\ph{\eta} \downarrow} \phi_{1s}\delta_{K_1,0}\Big) \delta_{m,x}
	\nn
	&+ i \Big(\big(q_{\eta l K_1}^{\ph{\eta} \downarrow} \phi_{1s}\big)^* \delta_{K_2,0} 
	\! + \! q_{\eta l K_2}^{\ph{\eta} \downarrow} \phi_{1s}\delta_{K_1,0}\Big) \delta_{m,y} \bigg]
	- i \frac{I \Jpd}{\hbar} F_{\eta\ph{_e} 1s 1s}^{\etae K_1 K_2} \frac{1}{2} \Big( \langle S^zS^l \rangle s_{K_2}^m \! - \! \langle S^lS^z \rangle s_{K_1}^m \Big)
	\nn
	&+ i \frac{I \Jsd}{2 \hbar} F_{\ph{-}\eta\ph{_h} 1s 1s}^{-\etah K_1 K_2} \sum_j \Big( 
	\langle S^jS^l \rangle \big( \frac{1}{2}\delta_{j,m} n_{K_2} - i \sum_k \epsilon_{jkm} s_{K_2}^k \big) 
	- \langle S^lS^j \rangle \big( \frac{1}{2}\delta_{j,m} n_{K_1} + i \sum_k \epsilon_{jkm} s_{K_1}^k \big)\Big),
	\an
\label{eq:EOM-for-summed-variables-Z1}
\ddt Z_{\eta \ph{0} K_1}^{\ph{\eta} 0 K_2} =&
	- i \big(\omega_{K_2} - \omega_{K_1}\big) Z_{\eta \ph{0} K_1}^{\ph{\eta} 0 K_2}
	+ \frac{i}{2\hbar}\mbf E \cdot \mbf M \Big(\big(z_{\eta K_1}^{\ph{\eta} \uparrow} \phi_{1s}\big)^* \delta_{K_2,0} 
	- z_{\eta K_2}^{\ph{\eta} \uparrow} \phi_{1s}\delta_{K_1,0}\Big)
	\nn
	&+ i \frac{I}{\hbar} \Big( J_0^\t{e} F_{\ph{-}\eta\ph{_h} 1s 1s}^{-\etah K_1 K_2} 
	+ J_0^\t{h} F_{\eta\ph{_e} 1s 1s}^{\etae K_1 K_2} \Big) \big( n_{K_2} - n_{K_1} \big),
	\an
\label{eq:EOM-for-summed-variables-Z2}
\ddt Z_{\eta \ph{l} K_1}^{\ph{\eta} l K_2} =&
	- i \big(\omega_{K_2} - \omega_{K_1}\big) Z_{\eta \ph{l} K_1}^{\ph{\eta} l K_2}
	+ \frac{i}{2\hbar}\mbf E \cdot \mbf M \bigg[ \Big(\big(z_{\eta K_1}^{\ph{\eta} \uparrow} \phi_{1s}\big)^* \delta_{K_2,0} 
	- z_{\eta K_2}^{\ph{\eta} \uparrow} \phi_{1s}\delta_{K_1,0}\Big)\delta_{l,z}
	\nn
	&+ \Big(\big(z_{\eta K_1}^{\ph{\eta} \downarrow} \phi_{1s}\big)^* \delta_{K_2,0} 
	- z_{\eta K_2}^{\ph{\eta} \downarrow} \phi_{1s}\delta_{K_1,0}\Big)\delta_{l,x}
	+ i \Big(\big(z_{\eta K_1}^{\ph{\eta} \downarrow} \phi_{1s}\big)^* \delta_{K_2,0} 
	+ z_{\eta K_2}^{\ph{\eta} \downarrow} \phi_{1s}\delta_{K_1,0}\Big)\delta_{l,y} \bigg]
	\nn
	&+ i \frac{I}{\hbar} \Big( J_0^\t{e} F_{\ph{-}\eta\ph{_h} 1s 1s}^{-\etah K_1 K_2}
	+ J_0^\t{h} F_{\eta\ph{_e} 1s 1s}^{\etae K_1 K_2} \Big) \big( s_{K_2}^l - s_{K_1}^l \big),
\end{align}
\end{subequations}
\end{widetext}
where $\phi_{1s} := R_{1s}(r = 0)$ is the radial part of the $1s$ exciton wave function evaluated at $r = 0$.
The factor $I = 3/2$ stems from the projection onto the quantum well and the exciton form factors are given by
\begin{align}
\label{eq:angle-averaged-form-factors}
F_{\eta_1 1s 1s}^{\eta_2 \omega_1 \omega_2} =&\; 2\pi \! \int_0^{2\pi} \!\!\!\! d\psi \! \int_0^\infty \!\!\!\! dr \! \int_0^\infty \!\!\! dr' \, r r' R_{1s}^2(r) R_{1s}^2(r')
	\nn
	&\times J_0\big( \eta_1 K_{12}(\psi) r \big) J_0\big( \eta_2 K_{12}(\psi) r' \big)
\end{align}
with the cylindrical Bessel function of order zero $J_0(x)$ and $K_{12} = |\K_1 - \K_2|$ with angle between $\K_1$ and $\K_2$ given by $\psi$.

\section{Markovian equations of motion for the exciton ground state}
\label{app:B}

When the magnetic and nonmagnetic scattering of excitons with impurities as well as the optical excitation are treated as a Markovian processes, one obtains the following 
equations of motion for the spin-up and spin-down exciton densities, respectively \cite{Ungar_Quantum-kinetic_2017, Ungar_Trend-reversal_2018}:
\begin{widetext}
\begin{align}
\label{eq:Markov-limit}
\ddt n_{\omega_1}^\ud =&\;
	\Gamma_{\omega_1} + \frac{35 I \NMn M \Jsd^2}{12 \hbar^3 V d}
	F_{\etah 1s 1s}^{\etah \omega_1 \omega_1} (n_{\omega_1}^\du - n_{\omega_1}^\ud)
	\nn
	&+ \int_0^\infty D(\omega) \Lambda_{1s 1s}^{\omega_1 \omega} \bigg[ \Theta\big(\omega - \omega_1 - \omega_{\omega-\omega_1}^\t{ph}\big)
	\Big( n_{\omega}^\ud \big(1+n^\t{ph}(\omega - \omega_1)\big) - n_{\omega_1}^\ud n^\t{ph}(\omega - \omega_1) \Big)
	\nn
	&+ \Theta\big(\omega_1 - \omega - \omega_{\omega_1-\omega}^\t{ph}\big)
	\Big( n_{\omega}^\ud n^\t{ph}(\omega_1 - \omega) - n_{\omega_1}^\ud \big(1+n^\t{ph}(\omega_1 - \omega)\big) \Big)\bigg].
\end{align}
\end{widetext}
The equations are formulated in frequency space with the constant density of states $D(\omega) = \frac{VM}{2\pi\hbar d}$ to achieve a better numerical
evaluation of the delta functions.
The frequency is connected to the exciton wave number via $\omega = \frac{\hbar K^2}{2M}$.
Furthermore,
\begin{align}
\label{eq:Gamma}
\Gamma_{\omega_1}(t) &= \frac{1}{\hbar^2} E(t) E_0 |M_\ud|^2 \phi_{1s}^2 \int_{-\infty}^t d\tau e^{-\frac{\tau^2}{2\sigma^2}} \, \delta_{\omega_1,0}
\end{align}
is the optical generation rate of excitons with $\sigma$ related to the time $t_\t{FWHM}$ at full-width half-maximum of the pulse via $\sigma = \frac{t_\t{FWHM}}{2\sqrt{2\log 2}}$.
Note that $|M_\ud|^2$ still contains spin selection rules.
From the spin-up and spin-down exciton density, the $z$ component of the spin can be extracted via $s_\omega^z = \frac{1}{2} (n_\omega^\up - n_\omega^\down)$.

\bibliography{references}
\end{document}